\documentclass[aps,amsfonts,nofootinbib,superscriptaddress,preprint]{revtex4-1}
\usepackage[normalem]{ulem} 
\usepackage{tikz}
\usepackage{cancel}
\usepackage{tikz-feynman}
\usepackage{caption}
\usepackage{subcaption}
\usepackage{graphicx}
\makeatletter
\makeatletter
\usepackage{tikz}
\usetikzlibrary{decorations.pathmorphing, shapes.misc, shapes}
\usepackage{amsmath}
\usepackage{mdframed} 
\usepackage{tikz-feynman}
\tikzfeynmanset{compat=1.1.0}
\usepackage{tikz-feynman} 
\tikzfeynmanset{compat=1.1.0} 
\usepackage{feynmp-auto}
\usepackage{cleveref}
\usepackage{amssymb}
\usepackage{xcolor} 
\usepackage{tcolorbox} 
\usepackage{amsbsy}
\usepackage{bm}
\usepackage{epsfig} 
\usepackage{color}
\usepackage{slashed}
\usepackage{soul}
\usepackage{comment}
\usepackage{auncial}
\makeatletter
\usepackage{appendix}
\usepackage{epsfig} 
\usepackage{ulem}
\usepackage{pgfplots}
\newcommand{\stkout}[1]{\ifmmode\text{\sout{\ensuremath{#1}}}\else\sout{#1}\fi}
\newcommand{\ee}{\end{equation}}
\newcommand{\bb}{\begin{equation}}
\newcommand{\eqb}{\begin{eqnarray}}

\newcommand{\eqf}{\end{eqnarray}}

\usepackage[T1]{fontenc}
\usepackage{comment}
\usepackage[normalem]{ulem} 
\makeatletter
\usepackage{xcolor}
\usepackage{mdframed}
\usepackage{amsmath}
\usepackage{amssymb}
\usepackage{appendix}
\usepackage{amsbsy}
\usepackage{tikz}
\usepackage{booktabs}
\usetikzlibrary{decorations.pathmorphing, shapes.misc, shapes}
\usepackage{amsmath}
\usepackage{tikz}
\usepackage{feynmp-auto}
\usepackage{cleveref}
\usepackage{bm}
\usepackage{epsfig} 
\usepackage{color}
\usepackage{slashed}
\usepackage{soul}
\usepackage{comment}
\usepackage[normalem]{ulem} 
\makeatletter
\usepackage{amsmath}
\usepackage{amssymb}
\usepackage{appendix}
\usepackage{amsbsy}
\usepackage{bm}
\usepackage{epsfig} 
\usepackage{color}
\usepackage{slashed}
\usepackage{soul}
\usetikzlibrary{positioning,arrows,patterns,automata}
\usetikzlibrary{decorations}
\usetikzlibrary{trees}
\usetikzlibrary{decorations.pathmorphing}
\usetikzlibrary{decorations.markings}
\usetikzlibrary{calc}

\usepackage{feynmp-auto} 
\begin{document}
\title{Topology and the Infrared Structure of Quantum Electrodynamics }
\author{J. Gamboa
}
\email{jorge.gamboa@usach.cl}
\affiliation{Departamento de F\'{\i}sica, Universidad de Santiago de Chile, Santiago 9170020, Chile
}

\begin{abstract}
We study infrared divergences in quantum electrodynamics using geometric phases and the adiabatic approximation in quantum field theory. In this framework, the asymptotic \textit{in} and \textit{out} states are modified by Berry phases, $e^{i \Delta \alpha_{\text{in}}}$ and $e^{i \Delta \alpha_{\text{out}}}$, which encode the infrared structure non-perturbatively and regulate soft-photon divergences.

Unlike the Faddeev--Kulish formalism, which employs perturbative dressing with coherent states, our approach reformulates the effective action in terms of Berry connections in field space. This yields finite, gauge-invariant scattering amplitudes without requiring a sum over soft-photon emissions.

We show that infrared divergences cancel to all orders in the bremsstrahlung vertex function $\Gamma^\mu(p_1, p_2)$, due to destructive interference among inequivalent Berry phases. As an application, we study the formation of positronium in the infrared regime and argue that the dressed \(S\)-matrix exhibits a functional singularity at $s = 4 m_e^2$, corresponding to a physical pole generated by topological flux.

\end{abstract}
\maketitle

\section{Introduction}

In Quantum Electrodynamics (QED), the infrared problem refers to divergences that arise in scattering amplitudes due to the masslessness of the photon. 

In scattering processes involving charged particles, an arbitrary number of soft photons (\( \omega \to 0 \)) can be emitted. The differential probability for emitting a soft photon with energy \( \omega \) is proportional to \( d\omega/\omega \). Integrating over all possible soft photon energies down to an infrared cutoff leads to infrared divergences of the form:
\begin{equation}
\int^\Lambda_\epsilon \frac{d\omega}{\omega} \sim \ln\left( \frac{\Lambda}{\epsilon} \right), \label{eq1}
\end{equation}
where \( \epsilon \) is the infrared cutoff and \( \Lambda \) is an ultraviolet cutoff. As \( \epsilon \to 0 \), the probability for exclusive scattering processes (with no soft photon emission) diverges due to the emission of an infinite number of arbitrarily soft photons.

This phenomenon was first addressed by Bloch and Nordsieck~\cite{BlochNordsieck} and later rigorously formulated by Yennie et al.~\cite {YFS} and Kinoshita~\cite{Kinoshita} and Lee and Nauenberg~\cite{LeeNauenberg}. They demonstrated that while exclusive amplitudes are divergent, inclusive --summing over soft-photon emissions-- remain finite.

An essential feature of QED in the infrared regime is the fundamental inseparability between fields and interactions. Charged particles are never isolated entities; instead, they are always accompanied by long-range electromagnetic fields and an infinite cloud of soft photons \cite{KulishFaddeev,chung} (see also \cite{ref1,ref2}). This inseparability implies that the classical notion of a ``free particle'' moving without interaction becomes invalid at asymptotic times $t \to \pm \infty$) \cite{FrohlichStrocchi,Herdegen}.

In standard scattering theory (e.g., the LSZ formalism), one assumes the existence of asymptotic free states that evolve independently of interactions at large times. However, in QED, the photon is exactly massless; electromagnetic interactions are long-ranged and;  the charged particles continuously interact with the field, even asymptotically and, as a consequence: 1) the Fock space of free electrons and photons is inadequate to describe physical asymptotic states and, 2) Scattering amplitudes computed between standard Fock states suffer from infrared divergences and may even vanish.

Thus, the proper identification of initial and final physical states becomes a highly nontrivial aspect of the infrared structure of the theory.
This is not unique to QED: In any quantum field theory with significant long-range effects or strong non-perturbative dynamics, the construction of asymptotic states requires careful treatment \cite{FrohlichStrocchi}.

To our knowledge, the first to address this problem differently from the approaches known up to the 1960s were Kulish and Faddeev \cite{KulishFaddeev,chung}, who recognized the ``dressed'' character of asymptotic states through the introduction of the operator \( \mathcal{W}(A) \). This operator arises from the asymptotic structure of the interaction \( J_\mu A^\mu \) in QED, specifically by focusing on the effective fermionic current at large times. Their essential idea was to define an asymptotic fermionic current such that coherent clouds of soft photons are generated around electrons, ensuring a consistent infrared structure for the physical states.

Although the Kulish-Faddeev method \cite{KulishFaddeev,chung} offers an elegant resolution of the infrared problem, it exhibits two key limitations that constrain its broader applicability. First, the dressing operator \( \mathcal{W}(A) \) is built under the assumption that electrons propagate along asymptotically straight trajectories as \( t \to \pm \infty \). Second, the construction remains fundamentally perturbative, making it less suitable for addressing non-perturbative infrared dynamics or strongly coupled gauge theories.

Motivated by these limitations, we now turn to an alternative framework that reinterprets the infrared structure of QED through functional Berry phases \cite{Berry}. In this approach, infrared effects are absorbed into a geometric phase associated with configurations, providing a more flexible and potentially non-perturbative description of asymptotic states.

More precisely, in Section II we formulate the generating functional of QED following Fujikawa's method, and construct an effective action using the adiabatic approximation, introducing the notion of the adiabatic gauge. Section III discusses dressed states and explains in detail their role as physical regulators in terms of accumulated Berry phases; in section IV, we analyze positronium as an application of the formalism and demonstrate that it emerges as a non-perturbative bound state of the dressed S-matrix. Finally in section V discusses the results and their implications.

\section{QED in the Fujikawa formalism and Berry Phase}

In this section, we formulate QED using the functional integral approach in order to make both gauge symmetry and the breaking of local chiral symmetry through the anomaly explicit. Although QED is not usually studied in this manner, as we will see later, this perspective proves useful for analyzing the infrared behavior of the theory.

Let us begin by considering the following generating functional \cite{fujikawa}:
\begin{equation}
\mathcal{Z} = \int [\mathcal{D}A]\, \mathcal{D}\bar{\psi}\, \mathcal{D}\psi ~e^{-S}, \label{f1}
\end{equation}
where the Euclidean action is given by
\begin{equation}
S = \int d^4x \left[ \bar{\psi}(\slashed{D} - \slashed{\partial} \alpha\, \gamma_5)\psi + \frac{1}{4} F_{\mu\nu} F^{\mu\nu} + \frac{1}{16\pi^2} (\theta - \alpha)\, \tilde{F}_{\mu\nu} F^{\mu\nu} \right]. \label{f2}
\end{equation}
The \(\alpha\)-dependent terms arise from a local chiral transformation of the fermion fields:
\[
\psi' = e^{i \alpha(x) \gamma_5}~ \psi,~~~~{\bar \psi}'= {\bar \psi}~~e^{i \alpha(x) \gamma_5}\]
under which the classical Dirac action transforms covariantly. The final term in the action, proportional to \((\theta - \alpha)\), originates from the noninvariance of the fermionic path integral measure captured by the Fujikawa Jacobian and encodes the chiral anomaly.

As a tentative step, promoting \(\theta \to \theta(x)\) and assuming that a locally well-defined \(\alpha(x)\) can be constructed, one may consider extending the axion field to a broader class of backgrounds than those typically considered. However, this extension is not straightforward, and certain topological considerations must be taken into account.

In particular, if \(\alpha(x)\) cannot be defined globally due to topological obstructions (e.g. winding or nontrivial bundles), the term \((\theta(x) - \alpha(x))\, \tilde{F} F\) must be interpreted carefully. In such cases, \(\alpha(x)\) may only exist locally in patches, with transition functions encoding a non-trivial gauge structure. The combination \((\theta (x) - \alpha)\, \tilde{F} F\) can then be understood as a globally defined object constructed from these local data, analogous to the descent relations in anomaly inflow or Wess–Zumino–Witten-type terms \cite{WessZumino,WittenWZW}. Alternatively, this expression can be viewed as the boundary of a five-dimensional Chern–Simons-like term, ensuring that the global contribution to effective action remains well defined modulo \(2\pi\). This highlights the intrinsic topological nature of the anomaly and the physical relevance of the Berry bundle structure.

We have written the expression (\ref{f2}) in this form because it is well suited for applying the adiabatic approximation. In fact, upon integration of the fermions, the determinant can be expressed in terms of the eigenvalue equation:
\[\left (\slashed{D} - i \gamma_5\slashed{\partial} \alpha + \slashed{\cal A}\right)\varphi_n = \lambda_n \varphi_n.
\]
We solve this equation using the Berry ansatz and rewrite the action (\ref{f2}) and the fermionic part becomes \cite{Gamboa}
\[
\bar{\psi}(\slashed{D} - i \gamma_5 \slashed{\partial} \alpha + \slashed{\mathcal{A}})\psi,
\]
where \({\cal A}_\mu\) is the Berry connection.

The next step is to choose the adiabatic gauge,
\bb
\mathcal{A}_\mu = i\gamma_5 \partial_\mu \alpha, \label{gauge1}
\ee
which implies that the total accumulated phase \(\Delta \alpha\) is given by
\bb
\Delta \alpha = i\gamma_5 \int_{\mathcal{C}} \mathcal{A}_\mu\, dx^\mu. \label{cond1}
\ee

Although the Berry connection takes the local form $\mathcal{A}_\mu = i\gamma_5 \partial_\mu \alpha$, it should not be regarded as a pure gauge in the usual sense. This is because \(\mathcal{A}_\mu\) arises from parallel transport in the Berry bundle associated with the family of eigenstates of the Dirac operator in the presence of a slowly varying chiral rotation \(\alpha(x)\). The parameter \(\alpha(x)\) originates from a local chiral transformation, which connects physically distinct vacua due to the chiral anomaly. As a result, the Berry connection encodes the geometry of the quantum bundle over the space of background fields.

The condition (\ref{cond1}) 
implies that the phase \( \alpha(x) \) is only globally defined and encodes topological information that must be carefully analyzed. The term \( (\theta(x) - \alpha)\, \tilde{F}^{\mu \nu} F_{\mu \nu} \) is also topologically significant, but for the purposes of this work --namely, the infrared implications for soft photons-- no further topological analysis is required.
\section{Topological Dressed States}

We now discuss the implications for physical observables, taking into account the adiabatic approximation and the Berry phase. As discussed above, in the adiabatic gauge the phase \( \alpha(x) \) cannot be defined locally, and it is more convenient to dress the asymptotic states with the accumulated Berry phase \( \Delta \alpha \). Thus, the in and out states become:
\begin{equation}
|i\rangle \to e^{i \Delta \alpha_{\text{in}}} |i\rangle, \qquad |f\rangle \to e^{i \Delta \alpha_{\text{out}}} |f\rangle, \label{modi1}
\end{equation}
implying that the physical amplitude \( \mathcal{M}_{\text{phys}} \) should be understood as:
\begin{equation}
\mathcal{M}_{\text{phys}} = e^{i(\Delta \alpha_{\text{in}} - \Delta \alpha_{\text{out}})} \mathcal{M}_0, \label{ampl1}
\end{equation}
where \( \mathcal{M}_0 \) is the amplitude computed between undressed (bare) in and out states.

The dressing of states is reminiscent of the Kulish–Faddeev approach \cite{KulishFaddeev,chung}, but as emphasized in the introduction, it differs in several important aspects; our formulation is inherently topological and non-perturbative.

Equation~(\ref{modi1}) may appear naively trivial, since the accumulated phases cancel when computing observables proportional to \( |\mathcal{M}_{\text{phys}}|^2 \), such as the cross section. However, in any realistic situation, soft photons with arbitrarily low energy cannot be individually detected. Therefore, the physical probability must include a \emph{sum} (or functional integral) over all possible soft-photon configurations.

In this case, each soft-photon configuration corresponds to a different value of the accumulated phase \( \Delta \alpha_{\text{out}} \); the resulting phase differences between configurations lead to destructive interference. This interference modifies the total transition probability in a nontrivial way. As a result, summing over soft final states introduces a functional dependence on the Berry phases. Physically, this leads to the suppression of infrared divergences that would otherwise arise when summing over exclusive final states with different numbers of soft photons.

\subsection{ Infrared divergences and Bremsstrahlung}

To illustrate the ideas discussed above, consider the bremsstrahlung process\footnote{We denote by $p$ and $p'$ the initial and final momenta, respectively, in this reaction.}

\begin{equation}
e^- (p)\to e^-(p') + \gamma(k), \label{soft1}
\end{equation}
where \( \gamma(k) \) denotes a soft photon. At first glance, this process appears problematic, as the corresponding emission rate becomes ill-defined in the limit \( \omega \to 0 \). Although the amplitude itself remains finite, the inclusive probability diverges logarithmically due to the integration over soft photon modes (see eq. (\ref{eq1}). 
                                                                      
This infrared divergence signals the breakdown of the exclusive description of scattering in QED and highlights the need for an appropriate infrared regulator or a redefinition of physical states.

For the process (\ref{soft1}) the tree-level amplitude is:
\begin{equation}
\mathcal{M} = e\, \epsilon_\mu(k) \left( \frac{p^\mu}{p \cdot k} - \frac{p'^\mu}{p' \cdot k} \right) \mathcal{M}_0.
\end{equation}

The rate --it is the probability per unit time that a soft photon is emitted-- involves the integral:
\begin{equation}
\Gamma_{\text{soft}} \sim \int_\epsilon^\Lambda \frac{d^3k}{(2\pi)^3 2\omega} \left| \frac{p}{p \cdot k} - \frac{p'}{p' \cdot k} \right|^2 \sim \ln\left( \frac{\Lambda}{\epsilon} \right), \label{9}
\end{equation}
which shows a logarithmic infrared divergence as \( \epsilon \to 0 \). Equation (\ref{9}) is universal and depends only on the kinematics, not on the details of the process.

One possible way to avoid infrared divergence is to sum over all physically indistinguishable final states; this is the original solution of Bloch and Nordsieck \cite{BlochNordsieck}, later improved and rigorously formulated in \cite{YFS}.

However, there is another, different, yet equivalent way to cancel the infrared divergence by invoking geometric arguments, namely, by using the idea of Berry phases, as we discussed above.

{{Indeed, let us analyze the same scattering process using the geometric approach introduced in this work. In this framework, the infrared (IR) structure of QED is treated non-perturbatively by dressing the asymptotic states with a Berry phase, accumulated due to the adiabatic evolution of the gauge background. This procedure is equivalent to redefining the standard Fock-space in/out states as dressed states that include the effect of an infinite number of soft photon modes.}}

{{To implement this idea, we perform a field-dependent phase rotation of the fermionic states, as encoded in Eq.~(\ref{modi1}). This transformation induces a geometric phase ---the Berry phase--- which modifies the asymptotic states. Thus, instead of computing the scattering amplitude between bare in/out states, we evaluate the transition amplitude between states dressed by these Berry phases. The amplitude then takes the form}}
{{
\bb
\mathcal{M}_{\text{phys}} = e^{i(\Delta \alpha_{\text{in}} - \Delta \alpha_{\text{out}})} \mathcal{M}_0, \label{amp01}
\ee}}
{{where \( \Delta \alpha_{\text{in}} \) and \( \Delta \alpha_{\text{out}} \) denote the Berry phases associated with the adiabatic evolution of the incoming and outgoing states, respectively, and \( \mathcal{M}_0 \) is the standard QED amplitude computed in perturbation theory.}}

While each individual Berry phase \( \Delta \alpha \) may be complex, the observable quantity is the squared modulus of the physical amplitude:
\begin{equation}
|\mathcal{M}_{\text{phys}}|^2 = |\mathcal{M}_0|^2 \cdot \left| e^{i(\Delta \alpha_{\text{in}} - \Delta \alpha_{\text{out}})} \right|^2 = |\mathcal{M}_0|^2 \cdot e^{-2\, \text{Im}(\Delta \alpha_{\text{in}} - \Delta \alpha_{\text{out}})}. 
\label{ima}
\end{equation}
{{This means that only the imaginary part of the Berry phase difference contributes to the physical probability. In our formalism, this imaginary component arises intrinsically from the adiabatic approximation, rather than from an explicit integration over soft photon modes. It reflects the loss of quantum coherence associated with unobservable infrared radiation, and leads to a real and non-local functional of the asymptotic particle trajectories.

This functional yields a universal and negative suppression factor in the squared amplitude, replacing the infrared divergences that would otherwise arise in perturbation theory. We emphasize that the accumulated Berry phase \( \Delta \alpha \) is not necessarily real: its imaginary part encodes the infrared suppression factor according to
\bb
\beta(p, p') = 2\, \text{Im}(\Delta \alpha(p) - \Delta \alpha(p')). \label{numer001}
\ee
}}

{{More explicitly, we find that the imaginary part of the Berry phase difference is given by the IR-divergent integral
\bb
\beta(p, p') = \frac{e^2}{8\pi^2} \int \frac{d^3k}{(2\pi)^3 2\omega_k} \left( \frac{p}{p \cdot k} - \frac{p'}{p' \cdot k} \right)^2, \label{numer01}
\ee
where \( p \) and \( p' \) are the four-momenta of the incoming and outgoing electrons, respectively, and the integrand reflects the emission and absorption of soft photons along each asymptotic trajectory. This expression is equivalent to the eikonal integral that appears in the exponentiation of IR divergences in conventional QED.}}

{{Therefore, the physical amplitude becomes
\bb
|\mathcal{M}_{\text{phys}}|^2 = |\mathcal{M}_0|^2 \cdot \left( \frac{\epsilon}{\Lambda} \right)^{-2\beta(p,p')}, \label{numer1}
\ee
where \( \epsilon \) is an IR regulator (e.g., photon mass or energy resolution scale), and \( \Lambda \) is a high-energy reference scale. This result shows that the IR divergence that arises in the perturbative calculation of \( |\mathcal{M}_0|^2 \) is precisely canceled by the geometric dressing factor. The exponent \( \beta(p, p') \) plays the same role as in the classic Bloch--Nordsieck or Kulish--Faddeev analyses, but here it arises naturally from the geometric structure of the adiabatic phase space.}}

This result reproduces, from a geometric and functional perspective, the known infrared structure of QED amplitudes, first explained by Bloch and Nordsieck \cite{BlochNordsieck} and later formalized by Kulish and Faddeev \cite{KulishFaddeev,chung} through a coherent dressing of charged states. 

In our approach, the exponential suppression emerges naturally from the interference between Berry phases associated to different asymptotic configurations, providing a non-perturbative resummation of soft-photon effects and a consistent framework for understanding infrared-finite amplitudes without explicit summation over soft photon states.

\subsection{Infrared Divergences at One Loop}

The one-loop vertex correction in QED is given by:
\begin{equation}
\Gamma^\mu(p',p) = (-ie)^2 \int \frac{d^4k}{(2\pi)^4} \gamma^\rho \frac{(\slashed{p}' - \slashed{k} + m)}{(p'-k)^2 - m^2 + i\epsilon} \gamma^\mu \frac{(\slashed{p} - \slashed{k} + m)}{(p-k)^2 - m^2 + i\epsilon} \gamma_\rho \frac{1}{k^2 + i\epsilon}.
\end{equation}

In the infrared region \( k \to 0 \), the electron propagators can be approximated as:
\begin{align}
\frac{(\slashed{p}' - \slashed{k} + m)}{(p'-k)^2 - m^2 + i\epsilon} &\approx \frac{(\slashed{p}' + m)}{2p'\cdot k + i\epsilon}, \\
\frac{(\slashed{p} - \slashed{k} + m)}{(p-k)^2 - m^2 + i\epsilon} &\approx \frac{(\slashed{p} + m)}{2p\cdot k + i\epsilon}.
\end{align}

The numerator then factorizes, at leading order, in terms of \( \gamma^\mu \). The diagram becomes dominated by:
\begin{equation}
\Gamma^\mu_{\text{IR}}(p',p) \approx 2(-ie)^2 \left( 2 {p'}^\mu \slashed{p} - \slashed{p}'\gamma^\mu \slashed{p}\right) \cdot I_{\text{IR}}, 
\end{equation}
with 
\[
I_{\text{IR}}=\int \frac{d^4k}{(2\pi)^4} \frac{1}{(2p'\cdot k)
(2p'\cdot k)(k^2)}
\]

This integral is well known and standard and, when evaluated with an infrared cut-off \( \epsilon \), the result is:
\[
I_{\text{IR}} \sim \frac{1}{(4\pi)^2} \cdot \frac{1}{p \cdot p'} \cdot \ln\left( \frac{\Lambda}{\epsilon} \right).
\]

The leading infrared divergence is then:
\begin{equation}
\Gamma^\mu_{\text{IR}}(p',p) \sim \gamma^\mu \, \delta_{\text{IR}},
\end{equation}
with
\begin{equation}
\delta_{\text{IR}} = \frac{e^2}{8\pi^2} \ln\left( \frac{\Lambda}{\epsilon} \right).
\end{equation}

The one-loop corrected scattering amplitude takes the form:
\begin{equation}
\mathcal{M} \sim \mathcal{M}_0 (1 + \delta_{\text{IR}}),
\end{equation}
where \( \mathcal{M}_0 \) is the tree-level amplitude.

\subsection{Cancellation of Infrared Divergences and Berry Phases}

Now, in our approach based on Berry phases, each physical state is dressed with a functional phase $|\Psi\rangle \to e^{i \Delta \alpha} |\Psi\rangle$. Thus, the accumulated phase difference between initial and final states is:
\begin{equation}
\Delta (\Delta \alpha) = e^2 \int \frac{d^3k}{(2\pi)^3 2\omega_k} \left( \frac{p}{p\cdot k} - \frac{p'}{p'\cdot k} \right)^2.
\end{equation}

Evaluating this integral in the infrared region yields:
\begin{equation}
\Delta (\Delta \alpha) = \frac{e^2}{8\pi^2} \ln\left( \frac{\Lambda}{\epsilon} \right).
\end{equation}

Therefore, the accumulated Berry phase contributes a real suppression factor in the squared amplitude:
\begin{equation}
e^{- \Delta(\Delta \alpha)} \approx 1 - \frac{e^2}{4\pi^2} \ln\left( \frac{\Lambda}{\epsilon} \right).
\end{equation}

On the other hand, the bare amplitude receives an infrared divergent correction from virtual photons:
\begin{equation}
|\mathcal{M}_{\text{bare}}|^2 = |\mathcal{M}_0|^2 \left( 1 + \frac{e^2}{4\pi^2} \ln\left( \frac{\Lambda}{\epsilon} \right) \right).
\end{equation}

Multiplying both contributions, we obtain the physical amplitude squared:
\begin{equation}
|\mathcal{M}_{\text{phys}}|^2 = |\mathcal{M}_0|^2 \left( 1 + \frac{e^2}{4\pi^2} \ln\left( \frac{\Lambda}{\epsilon} \right) \right)
\left( 1 - \frac{e^2}{4\pi^2} \ln\left( \frac{\Lambda}{\epsilon} \right) \right)
+ \mathcal{O}(e^5).
\end{equation}

Expanding to first order:
\begin{equation}
|\mathcal{M}_{\text{phys}}|^2 = |\mathcal{M}_0|^2 \left( 1 + \cancel{\frac{e^2}{4\pi^2} \ln\left( \frac{\Lambda}{\epsilon} \right)} - \cancel{\frac{e^2}{4\pi^2} \ln\left( \frac{\Lambda}{\epsilon} \right)} \right)
+ \cdots ,
\end{equation}
where $\cdots$ denotes the term $\left(\frac{e^2}{4\pi^2}\right)^2 \ln^2 \left(\frac{\Lambda}{\epsilon}\right)$, which is the leading contribution of a divergent series that is, nonetheless, summable—as will be demonstrated.

This yields:
\begin{equation}
|\mathcal{M}_{\text{phys}}|^2 = |\mathcal{M}_0|^2 + \mathcal{O}(e^4).
\end{equation}

A useful technical remark is to note that the usual expression for the vertex
\begin{equation}
\Gamma^\mu(p',p) = \gamma^\mu F_1(q^2) + \frac{i \sigma^{\mu\nu} q_\nu}{2m} F_2(q^2),
\end{equation}
where \( q = p' - p \) and $F_1(q^2)$ and $F_2(q^2)$ are the form factors. 

In the adiabatic approximation, the effect of soft gauge fields is described by a geometric Berry phase \( \Delta\alpha \) that accumulates during the slow evolution of the system. The multiplicative phase factor $\Delta \alpha$ can be interpreted as an effective infrared-dependent modification of the electron interaction with the gauge field through a vertex correction. In this sense, the functional Berry phase captures the same infrared structure that is traditionally described by the form factor \( F_1(q^2) \), but from a geometric and dynamical perspective.

Thus, in the adiabatic framework, one can interpret the IR behavior of the vertex as emerging from the coherent dressing of the electron via \( \Delta\alpha \), which acts as an effective form factor in the soft limit.

\subsection{Infrared Divergence Cancellation Beyond Leading Order}

The arguments from the previous section can be extended more or less directly beyond leading order. In two loops, there are seven contributing diagrams, as shown in Figure~1. We focus exclusively on diagram (a) in Figure 1, since the formal structure is analogous to the remaining diagrams, namely \cite{Bonciani}

\eqb
&& \sim \int \frac{d^4k_1}{(2\pi)^4} \frac{d^4k_2}{(2\pi)^4}  
 \frac{{\cal N}^\mu }{
D_1 D_2 D_9 D_{10} D_{12} D_{13}
}, \label{13}
\eqf
where 
\eqb 
&&
{\cal N}^\mu= 
\bar{u}(p') \gamma^\sigma \left[ i(\slashed{p}' + \slashed{k}_1) \right] 
\gamma_\lambda \left[ -i(\slashed{p}' +\slashed k_1- \slashed{k}_2) \right] 
\gamma^\mu \left[ -i(\slashed{p} - \slashed{k}_1) +\slashed{k}_2 \right]\gamma^\lambda \left[-(\slashed{p}-\slashed{k}_1\right]
\gamma_\sigma u(p), \nonumber \\ \label{N1} 
\\
&&D_1= k_1^2, ~~~~D_2= k_2^2, ~~~~D_9= -2 p \cdot k_1, \nonumber
\\
&& D_{10}= 2 p' \cdot k_1,~~~~D_{12}= 2 p\cdot (k_2-k_1),~~~D_{13}= -2p'\cdot (k_2-k_1).\nonumber
\eqf

Using the definitions introduced in Eq.~(\ref{13}), we obtain
\bb
 \sim e^5 \int \frac{d^4k_1}{(2\pi)^4} \frac{d^4k_2}{(2\pi)^4} \, 
\frac{\bar{u}(p') \, \mathcal{N}^\mu(k_1, k_2; p,p') \, u(p)}
{k_1^2 \, k_2^2 \, (-2 p \cdot k_1)(-2 p' \cdot k_2)(-2 p' 
\cdot (k_1 + k_2))}.
\label{eq:vertex2loop}
\ee

In the infrared regime, the numerator $\mathcal{N}^\mu$ can be factorized and interpreted as a multiplicative correction. This structure reflects the accumulation of a Berry-type phase during the adiabatic evolution of the electron in a soft background field, if we denotes the accumulated Berry corrections of the vertex by $\delta\Gamma^\mu_{IR}$,  we obtain:

\begin{equation}
\delta \Gamma^\mu_{\text{IR}}(p, p') \approx \Gamma_0^\mu(p, p') \cdot \Delta \alpha^{(2)}(p, p'),
\end{equation}

where the two-loop $\Delta \alpha^{(2)}$ is given by:

\begin{equation}
\Delta \alpha^{(2)}(p, p') = e^2 \int \frac{d^4k_1}{(2\pi)^4} \frac{d^4k_2}{(2\pi)^4} \, 
\frac{F(p,p';k_1,k_2)}{k_1^2 \, k_2^2 \, ( 2p \cdot k_1)(p' \cdot k_2)(p' \cdot (k_1 + k_2))}.
\label{eq:berry2loop}
\end{equation}
where we can factor out the tree-level vertex and write
\[
\mathcal{N}^\mu(p,p',k_1, k_2) = \Gamma^\mu_0(p, p') \cdot F(p, p', k_1, k_2),
\]
with
\bb
F(p, p', k_1, k_2) = \left( \frac{p^\alpha}{p \cdot k_1} - \frac{{p'}^\alpha}{p' \cdot k_1} \right)
\left( \frac{p_{\alpha}}{p \cdot k_2} - \frac{p'_{\alpha}}{p' \cdot k_2} \right). \label{fact1}
\ee

The two-loop correction to the electron vertex finally becomes
\begin{equation}
\delta \Gamma^\mu_{\text{IR}}(p, p') \sim e^5 \int \frac{d^4k_1}{(2\pi)^4} \frac{d^4k_2}{(2\pi)^4} \, 
\frac{\bar{u}(p') \, \mathcal{N}^\mu(k_1, k_2) \, u(p)}
{k_1^2 \, k_2^2 \, (-2 p \cdot k_1)(2 p' \cdot k_2)(-2 p' \cdot (k_1 + k_2))}.
\label{eq:vertex2loop}
\end{equation}
Evaluating this expression in the limit of soft photon momenta $k_1, k_2 \ll m$, and applying an infrared cutoff $\epsilon$ and an ultraviolet cutoff $\Lambda$, the result contains a double logarithmic divergence:
\begin{equation}
\delta \Gamma^\mu_{\text{IR}}(p, p') \sim \Gamma^\mu_0(p, p') \cdot \left( \frac{e^2}{4\pi^2} \right)^2 \ln^2 \left( \frac{\Lambda}{\epsilon} \right).
\label{eq:vertex-divergence}
\end{equation}

and in ~\eqref{eq:vertex-divergence}, $\Gamma^\mu_0(p, p')$ denotes the tree-level QED vertex function, i.e.,
\bb
\Gamma^\mu_0(p, p') = \gamma^\mu, \label{gamma0}
\ee
which corresponds to the lowest-order interaction between a fermion and the gauge field. The correction \( \delta \Gamma^\mu_{\text{IR}} \) captures the leading infrared-divergent contribution at two loops, and is proportional to the tree-level structure times a double logarithmic enhancement in the infrared cutoff ratio.

However, this divergence does not appear in the physical amplitude, which includes adiabatic corrections to the external states. These corrections correspond to Berry phases accumulated by the incoming and outgoing electron states due to the soft gauge field background. The two-loop Berry phase correction is given by:
\begin{equation}
\Delta \alpha^{(2)}_{\text{in}} + \Delta \alpha^{(2)}_{\text{out}} \sim - \left( \frac{e^2}{4\pi^2} \right)^2 \ln^2 \left( \frac{\Lambda}{\epsilon} \right),
\label{eq:berry-cancellation}
\end{equation}
with a minus sign that exactly cancels the vertex divergence in Eq.~\eqref{eq:vertex-divergence}.

Therefore, the total infrared contribution to the physical amplitude at two loops becomes finite:
\begin{equation}
\mathcal{M}_{\text{phys}} = \left( \Gamma^\mu_0 + \delta \Gamma^\mu_{\text{IR}} \right) \cdot e^{i (\Delta \alpha_{\text{in}} + \Delta \alpha_{\text{out}})}                                 
\sim \Gamma^\mu_0,
\end{equation}
up to terms of order $e^5$ that are free of infrared divergences. This explicitly shows that the logarithmic divergence $\ln^2(\Lambda/\epsilon)$ is an unphysical artifact of perturbation theory and is canceled by the geometric phase structure of the soft photon sector.

The three-loop calculations above yield
\begin{align*}
\sim \int \frac{d^4 k_1}{(2\pi)^4} &\frac{d^4 k_2}{(2\pi)^4} \frac{d^4 k_3}{(2\pi)^4} 
\frac{{\cal N}^\mu (k_1,k_2,k_3; p,p')}{
k_1^2 \, k_2^2 \, k_3^2 \, 
(-2 p\cdot k_1) \, (-2 p \cdot (k_1+k_2))} \times \\
&\times \frac{1}{
(-2 p \cdot k_3) \, 
(-2 p \cdot (k_1+k_2+k_3)) \, 
(-2p' \cdot (k_1 + k_2)) \, 
(-2p' \cdot (k_1 + k_2 + k_3))} \\
&= \frac{1}{6} \, \beta^3 \left[ \ln \left( \frac{\Lambda}{\epsilon} \right) \right]^3.
\end{align*}

Examining each order of the perturbative series, the $n$-th term behaves as $(\ln x)^n / n!$, so the full perturbative series, although formally divergent, is actually summable. In fact,
\bb
\sim \sum_{n=0}^\infty \frac{1}{n!} \left( \beta \ln \frac{\Lambda}{\epsilon} \right)^n = \left( \frac{\Lambda}{\epsilon} \right)^\beta. \label{suda}
\ee

   The perturbative expansion of the amplitude squared yields the Sudakov factor:
\[
|\mathcal{M}|^2_{\text{pert}} = |\mathcal{M}_0|^2 \cdot \left( \frac{\Lambda}{\epsilon} \right)^{2 \beta(p, p')},
\]
which contains the infrared divergence as $\epsilon \to 0$.

However, in our formulation, the physical in and out states carry accumulated Berry phases:
\[
\Delta \alpha_{\text{in}} = \beta(p, p') \ln \left( \frac{\Lambda}{\epsilon} \right), \quad
\Delta \alpha_{\text{out}} = \beta(p, p') \ln \left( \frac{\Lambda}{\epsilon} \right),
\]
and thus introduce a correction factor:
\[
\left( \frac{\epsilon}{\Lambda} \right)^{2 \beta(p, p')}.
\]

Multiplying both contributions yields
\[
|\mathcal{M}_{\text{phys}}|^2 =
|\mathcal{M}_0|^2 \cdot 
\left( \frac{\Lambda}{\epsilon} \right)^{2 \beta(p, p')} \cdot
\left( \frac{\epsilon}{\Lambda} \right)^{2 \beta(p, p')} 
= |\mathcal{M}_0|^2,
\]
so the infrared divergences cancel exactly.

\begin{figure}
    \centering
    \begin{subfigure}[t]{0.35\textwidth}
       \includegraphics[width=\linewidth]{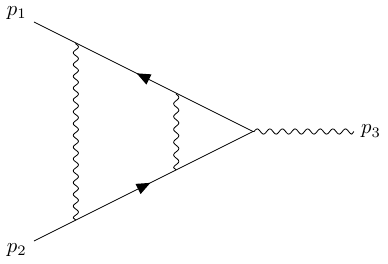} 
        \caption{} \label{fig:feyna}
           \end{subfigure}
    \begin{subfigure}[t]{0.35\textwidth}
        \centering
      \includegraphics[width=\linewidth]{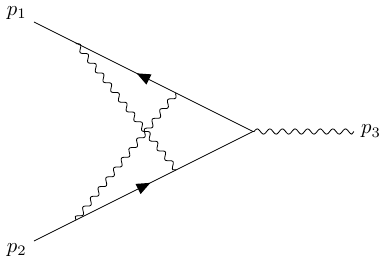} 
        \caption{} \label{fig:feynb}
    \end{subfigure}
    \vspace{1cm}
    \begin{subfigure}[t]{0.35\textwidth}
    \centering
        \includegraphics[width=\linewidth]{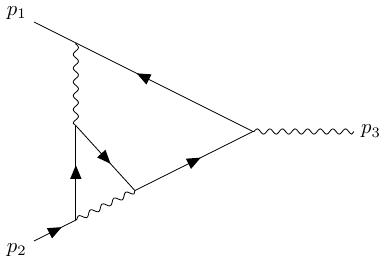} 
        \caption{} \label{fig:feync}
    \end{subfigure}
        \begin{subfigure}[t]{0.35\textwidth}
    \centering
        \includegraphics[width=\linewidth]{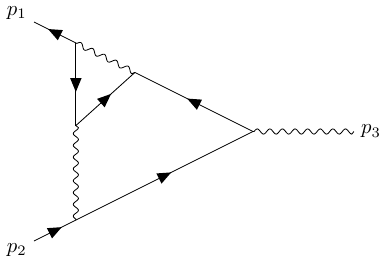} 
        \caption{} \label{fig:feynd}
    \end{subfigure}
        \begin{subfigure}[t]{0.35\textwidth}
    \centering
        \includegraphics[width=\linewidth]{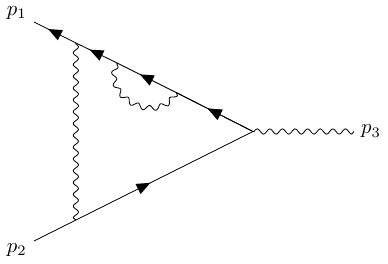} 
        \caption{} \label{fig:feyne}
    \end{subfigure}
        \begin{subfigure}[t]{0.35\textwidth}
    \centering
        \includegraphics[width=\linewidth]{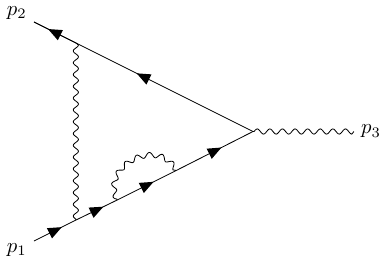} 
        \caption{} \label{fig:feynf}
    \end{subfigure}
        \begin{subfigure}[t]{0.35\textwidth}
    \centering
        \includegraphics[width=\linewidth]{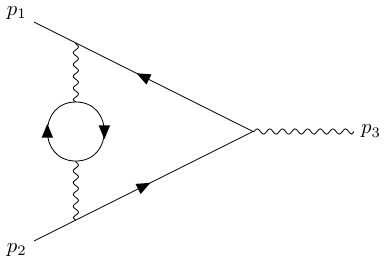} 
        \caption{} \label{fig:feung}
    \end{subfigure}
    \caption{Two loops diagrams}
\end{figure}

\section{Positronium}

As an application of the formalism discussed above, let us consider the problem of a bound electron-positron state, which by its very nature lies beyond the scope of conventional relativistic quantum field theory. This example is particularly interesting because, in this context, the asymptotic \textit{in} and \textit{out} states are dressed and carry non-trivial topological information at infinity.

The relevant question is therefore the following, what is the probability for positronium formation in the infrared regime of QED?. 

To address this problem, let us consider the $e^- e^+$ process and express the accumulated functional phase $\Delta \alpha(s)$ in the adiabatic gauge as

\begin{equation}
\Delta \alpha(s) = i \int dx^\mu\, \gamma_5\, \mathcal{A}_\mu[x(s)] 
= i \int dx^\mu\, \partial_\mu \alpha(x(s)) 
= i\left( \alpha(x_f(s)) - \alpha(x_i(s)) \right).
\end{equation}

Thus, if $\alpha$ is a multivalued function -as we assume here - then the resulting phase is nontrivial: $\Delta \alpha \neq 0$.

To address this problem, let us consider the $e^- e^+$ process and express the accumulated functional phase $\Delta \alpha(s)$ in the adiabatic gauge as
\begin{equation}
\Delta \alpha(s) = i \int dx^\mu\, \gamma_5\, \mathcal{A}_\mu[x(s)] 
= i \int dx^\mu\, \partial_\mu \alpha(x(s)) 
= i \left( \alpha(x_f(s)) - \alpha(x_i(s)) \right).
\end{equation}

If $\alpha$ is a multivalued function—as we assume here—then $\Delta \alpha \neq 0$. In this case, it can be represented by a logarithm, as this is the only analytic function with additive monodromy. In terms of the kinematical invariants of the $e^- e^+$ system, we write:
\begin{equation}
\Delta \alpha(s) = \gamma \ln \left( \frac{s - M^2 + i \epsilon}{\mu^2} \right),
\end{equation}
where $s = (p_{e^-} + p_{e^+})^2$.

By choosing the parameter $\gamma = 1$, as in the case of a magnetic monopole, the phase becomes quantized and the multivalued nature of $\Delta \alpha$ is made explicit. Since physical scattering processes are described by the dressed $S$-matrix, defined as
\begin{equation}
S^{\text{eff}}(s) = S(s) \cdot e^{i \Delta \alpha(s)},
\end{equation}
it is this effective matrix that captures the relevant infrared dynamics. The singular structure of $S^{\text{eff}}(s)$ at $s = M^2$ therefore reflects the presence of a bound state, generated nonperturbatively through the topological properties of the functional phase.

This behavior is entirely nonperturbative and emerges from the topological structure of the functional space of gauge fields. In this sense, the bound state appears not as a resummation of Feynman diagrams, but as a consequence of the monodromy of the Berry phase accumulated along a nontrivial loop in configuration space.

\section{Discussion of the Results}

We have shown that, in the infrared limit of QED—interpreted here as the analog of the adiabatic approximation—accumulated Berry phases naturally redefine the initial and final states. These phases act as finite and physically meaningful regulators of infrared divergences. 

This mechanism suggests that the infrared dynamics of QED is fundamentally geometric and topological in nature, and therefore intrinsically non-perturbative. In this framework, the infrared vacuum is not unique, but rather is labeled by non-trivial global data. A careful investigation of this structure may lead to previously unexplored physical consequences, possibly involving a deeper understanding of asymptotic states, vacuum sectors, and topological aspects of gauge theories.

However, in the case of QED, one may argue that the dressing of asymptotic states makes it possible to address bound-state formation, and that the topological mechanism associated with Berry phases provides a concrete route to a problem that has historically remained underexplored in relativistic quantum field theory\footnote{It is worth noting that Bethe and Salpeter \ first addressed the bound-state problem cite{BetheSalpeter1951,RobertsSchmidt2000} although their method suffers from limitations regarding covariance and relativistic invariance.}

Our approach features a non-perturbative vacuum structure, which we may refer to as a ``charged vacuum,'' reminiscent of the framework proposed by Fröhlich, Morchio, and Strocchi \cite{FrohlichStrocchi}. As in any effective theory of the Born–Oppenheimer type, this vacuum retains a form of ``memory''~ of the asymptotic configuration, and in this sense it resembles the infrared description discussed by Strominger \cite{Strominger}. 

Finally, we would like to briefly comment on some similarities and differences between our approach and the asymptotic analysis of infrared dynamics developed by Strominger and collaborators~\cite{strominger2014,strominger_memory,strominger_qed}. Both frameworks emphasize that infrared divergences in QED are not merely a technical issue, but reflect a deeper restructuring of asymptotic states.

In the asymptotic approach, soft modes and large gauge transformations are defined at future and past null infinity, $\mathcal{I}^\pm$, and the resulting memory effect appears as a permanent change in the electric field measured on the celestial sphere $S^2$. This structure is encoded in the angular dependence of large gauge parameters $\varepsilon(\hat{\mathbf{n}})$, which act non-trivially on physical states.

In contrast, our method operates entirely in the bulk. It uses the accumulated Berry phase $\Delta \alpha = \int \mathcal{A}$ -- evaluated in the adiabatic gauge $\mathcal{A}_\mu = i \gamma_5 \partial_\mu \alpha(x)$ -- to dress asymptotic states in a non-perturbative and gauge-invariant manner. Although $\mathcal{I}^\pm$ does not appear explicitly, the angular structure of the memory effect reemerges through the asymptotic behavior $\alpha(x) \to \alpha(\hat{\mathbf{n}})$ as $r \to \infty$.

A more detailed comparison, including a potential dictionary between asymptotic charges and Berry phases, is currently under development.

\section*{Acknowledgments}

The author thanks Fernando Méndez for his valuable and insightful discussions throughout this work. This research was supported by DICYT (USACH), grant number 042531GR$\_$REG. Finally, the author gratefully acknowledges the referee’s sharp and insightful comments, which helped to clarify important conceptual aspects.

\end{document}